\documentclass[nofootinbib,preprint,tightenlines,letterpaper,floatfix]{revtex4-1}

\usepackage{graphicx}
\usepackage{amsmath}
\usepackage{amssymb}

\newcommand{\e}{\mathrm{e}}
\newcommand{\dd}{\mathrm{d}}
\newcommand{\mul}{\mathrm{mul}}

\newcommand{\halb}{\frac{1}{2}}
\newcommand{\lp}{{\ell^{\prime}}}

\renewcommand{\mp}{{m^{\prime}}}

\newcommand{\zdrei}{\mathbb{Z}^3}
\newcommand{\ergunit}{$/(ma^2)$}

\def\mqo2{{\!\!\!}}

\begin{document}


\title{Three particles in a finite volume:\\ The breakdown of spherical  symmetry}

\author{Simon Kreuzer}
\email{E-Mail: skreuzer@gwu.edu (corresponding author)}
\author{Harald W. Grie{\ss}hammer}
\email{E-Mail: hgrie@gwu.edu}
\affiliation{Institute for Nuclear Studies, Department of Physics,\\
The George Washington University, Washington DC 20052, USA}

\date{\today}

\begin{abstract}
  Lattice simulations of light nuclei necessarily take place in finite
  volumes, thus affecting their infrared properties. These effects can
  be addressed in a model-independent manner using Effective Field
  Theories. We study the model case of three identical bosons
  (mass~$m$) with resonant two-body interactions in a cubic box with
  periodic boundary conditions, which can also be generalized to the
  three-nucleon system in a straightforward manner. Our results allow
  for the removal of finite volume effects from lattice results as
  well as the determination of infinite volume scattering parameters
  from the volume dependence of the spectrum. We study the volume
  dependence of several states below the break-up threshold, spanning
  one order of magnitude in the binding energy in the infinite volume,
  for box side lengths~$L$ between the two-body scattering length~$a$
  and $L=0.25a$. For example, a state with a three-body energy of
  $-3/(ma^2)$ in the infinite volume has been shifted to $-10/(ma^2)$
  at $L=a$. Special emphasis is put on the consequences of the
  breakdown of spherical symmetry and several ways to perturbatively
  treat the ensuing partial wave admixtures. We find their
  contributions to be on the sub-percent level compared to the strong
  volume dependence of the S-wave component. For shallow bound states,
  we find a transition to boson-diboson scattering behavior when
  decreasing the size of the finite volume.
\end{abstract}

\maketitle

\section{Introduction}

Quantum chromodynamics (QCD) is the theory underlying strong
interactions. However, ab initio calculations of hadronic and nuclear
properties remain one of the largest theoretical challenges of the
Standard Model. Lattice simulations provide a numerical approach, but
do at present not usually operate at the physical point~(for reviews
on the lattice simulations of light nuclei see,
e.g.,~\cite{Beane:2008dv,Beane:2010em}). In a complementary approach,
Effective Field Theories (EFT) describe the effective degrees of
nuclear physics, namely nucleons and pions, and allow for accurate
calculations of low-energy observables with a direct link to QCD
through the symmetries of the theory, see
e.g.~\cite{Beane:2000fx,Bedaque:2002mn,Phillips:2002da,Epelbaum:2005pn,
  Epelbaum:2008ga,Machleidt:2011zz}.

In lattice simulations, the QCD path integral is evaluated in a
discretized Euclidean space-time using Monte-Carlo simulations. This
approach requires a large numerical effort, which in turn strongly
constrains the parameters of the simulation. In particular, the
considered system is necessarily placed in a finite volume. Present
day calculations use cubic boxes with periodic boundary conditions and
relatively small side lengths of a few~fm. Momentum quantization due
to the boundary conditions causes a shift of the finite-volume
spectrum relative to the infinite-volume
energies~\cite{Luscher:1985dn}. A model-independent determination of
this shift is necessary in order to extract physical observables from
lattice results. Calculations at large volumes are possible, but
rendered inefficient by the enormous numerical effort
necessary. Therefore, the use of well-known physics to perform the
extrapolation is warranted.

The volume dependence of the spectrum also
provides access to scattering parameters. Most prominently,
L{\"u}scher showed that infinite volume scattering phase shifts as
well as resonance properties are encoded in the finite-volume spectrum
of two-particle states~\cite{Luscher:1990ux,Luscher:1991cf}.  

The correlation function for the three-nucleon system in the triton
channel has been calculated in Lattice QCD
recently~\cite{Beane:2009gs}, but because of the relatively large
uncertainties no triton properties could be extracted.  With
quantitative lattice data on light nuclei and their scattering
properties within reach, L{\"u}scher's results need to be extended to
the three-body sector in order to understand the finite volume effects
in these results.

The desired incorporation of well-known physics, combined with the
demand of model independence and the ability to achieve a given level
of accuracy, are preconditions well met by the EFT approach. In the
nuclear sector, the pionless EFT, which is valid for processes with
typical momenta below the pion mass, has been successfully used to
describe the properties of light nuclei~(see, e.g.,
\cite{Beane:2000fx,Bedaque:2002mn,Epelbaum:2008ga} for reviews). To
leading order, the three-body sector is described by the
nucleon-nucleon scattering lengths in the $^1S_0$ and $^3S_1$ channels
and a Wigner SU(4) symmetric three-body
force~\cite{Efimov-70,Bedaque:1999ve,Griesshammer:2005ga}. 

The pionless EFT in the nucleonic sector can be seen as part of a
larger family of EFTs for systems with resonant two-body
interactions. They are characterized by the appearance of a two-body
scattering length~$a$ that is unnaturally large compared to the range
of interaction. Such systems display interesting universal
properties. If~$a$ is positive, two particles of mass~$m$ form a
shallow two-body bound state with binding energy~$1/(ma^2)$,
independent of the detailed mechanism generating the large scattering
length. For example, $^4$He atoms have the unnaturally large
scattering length 189~$a_0$, where $a_0$ is the Bohr radius. Using the
aforementioned relation to extract $a$ from the energy of the helium
dimer, on the other hand, yields a scattering length of 182~$a_0$,
deviating by only 3.7\%. In the pionless EFT, the shallow two-body
bound state is identified as the deuteron.

In the three-body system, the universal properties include the Efimov
effect~\cite{Efimov-70}. If at least two of the three particles have a
scattering length~$|a|$ that is large compared to the range of their
interaction, a sequence of three-body bound states exists. In the
limit of diverging scattering length, there are infinitely many
geometrically spaced bound states with an accumulation point at
threshold. This spectrum is the signature of the Efimov effect, namely
a discrete scaling symmetry whose consequences can be calculated in an
EFT for short-range interactions. Here, it appears because the
renormalization group flow of the three-body coupling is a limit
cycle~(see, e.g.,~\cite{Braaten:2004rn,Hammer:2010kp} for a
review). This is the case for the Wigner SU(4) symmetric three-body
force in pionless EFT~\cite{Bedaque:1999ve}, but also for the
three-body force needed to renormalize the simpler EFT of three
identical bosons~\cite{Bedaque:1998km}. In this work, we will
therefore study systems of three identical bosons inside a cubic box
with periodic boundary conditions using the EFT
framework. Transferring the methods developed in this publication to
the pionless EFT will give an understanding of the finite volume
corrections in lattice calculations of the three-nucleon system.

In this paper, we study the volume dependence of three-boson states
below the three-body break-up threshold inside a cubic box with
periodic boundary conditions. One of the authors performed similar
studies
before~\cite{Kreuzer:2008bi,Kreuzer:2009jp,Kreuzer:2010ti,phd}. We
developed the new framework presented in this paper for several
reasons. First, the new approach provides access to the energy region
above the boson-diboson break-up threshold. Second, it does not rely
on the use of basis functions but remains in close contact with the
discrete space of allowed momenta. It further allows for a faster
numerical implementation when higher partial waves are taken into
account. Thus, for the first time, we are able to study the size of
these higher partial wave contributions in a systematic manner as well
as how they can be included perturbatively.

The application of EFT to finite volumes, especially for
three-particle systems, has sparked interest in recent years.  The
volume dependence for three spin-1/2 fermions with perturbative
interactions in a box has been studied
previously~\cite{Luu:2008fg}. Epelbaum and collaborators have
calculated the energy of the triton in a finite volume by implementing
a discretized version of chiral EFT on a
lattice~\cite{Epelbaum:2009zsa}. Also, the triton has been considered
in pionless EFT in a nuclear lattice formalism but the volume
dependence was not investigated \cite{Borasoy:2005yc}. Other volume
shapes have also been studied, most prominently the harmonic
oscillator~\cite{Tolle:2010bq,Rotureau:2011vf}, which allows to
compute finite volume effects inside atomic traps.

This paper is organized as follows. In the next Section, we derive the
basic equations and sketch the ideas of our numerical
implementation. We furthermore present two different approaches to
include higher partial wave corrections perturbatively. In
Section~\ref{results}, we provide a detailed discussion of our
numerical results, including a study on the perturbative nature of
higher partial waves. Section~\ref{summary} summarizes our findings
and provides an outlook for future work.

\section{Framework}
In the following, we derive a set of coupled equations governing the
partial waves of the amplitude and explain how to prepare these
equations for a numerical implementation. In the last part of this
section, we show two ways to take higher partial waves into account
perturbatively.

\subsection{Sum equation for the amplitude}
The Lagrangian for three identical bosons interacting via short-range
forces can be written as~(see,
e.g., \cite{Bedaque:1998km,Braaten:2004rn})
\begin{equation}\label{eq:lagrangian}
  \mathcal{L} = \psi^\dagger\left(i\partial_t+\halb\nabla^2\right)\psi +
  \frac{g_2}{4}d^\dagger d -
  \frac{g_2}{4}\left(d^\dagger\psi^2+\mathrm{H.c.}\right) -
  \frac{g_3}{36} d^\dagger d \psi^\dagger \psi + \dots,
\end{equation}
where the dots indicate higher order terms of the effective
theory. The Lagrangian is formulated in terms of the boson
field~$\psi$ and a non-dynamical auxiliary field~$d$ with the quantum
numbers of two bosons. Units have been chosen such that $\hbar=m=1$,
where $m$ is the mass of a single boson.

The system of three bosons is assumed to be contained in a cubic box
with side length~$L$ and periodic boundary conditions. This leads to
quantized momenta~$\frac{2\pi}{L}\vec{n}$, $\vec{n}\in\zdrei$.  As a
consequence, the loop integrations of the infinite volume case are
replaced by sums. If loop sums are divergent, they are regulated by a
cutoff~$\Lambda$ similar to the infinite volume case.

While the finite volume modifies the infrared regime of the theory by
introducing the low-momentum scale~$2\pi / L$, the ultraviolet
behavior does not change. Therefore, its renormalization is the same
in the finite and infinite volume cases. Of course, this statement is
only valid as long as the infrared and ultraviolet regime of the
theory characterized by the scales~$2\pi / L$ and~$\Lambda$,
respectively, are well separated, i.e., $\Lambda L \gg 1$. We will
explicitly demonstrate that the numerical results in this publication
are indeed renormalized. The cutoff dependence in the two-body sector
can be removed completely by matching the coupling constant~$g_2$ to a
low-energy two-body observable, namely the two-body scattering
length~$a$ or, if applicable, the two-body binding energy. As the
cutoff is increased, the three-body coupling approaches a
renormalization group limit cycle, leading to a log-periodic
dependence of the coupling constant~$g_3$ on the cutoff~$\Lambda$. For
convenience, $g_3$ is often expressed in terms of a dimensionless
function~$H$ via $g_3 = -9g_2^2\,H(\Lambda)/\Lambda^2$. The phase of
the log-periodic dependence of~$H$ has to be fixed from a three-body
datum~\cite{Bedaque:1998km,Braaten:2004rn,Platter:2009gz,Hammer:2010kp}.

\begin{figure}[t]
  \centering
  \includegraphics*[width=.75\linewidth]{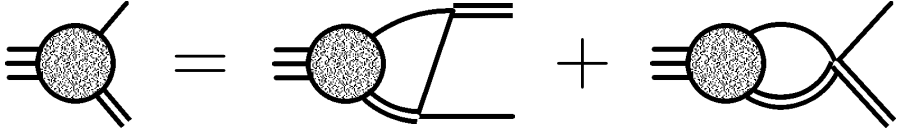}
  \caption{Integral equation for the boson-diboson amplitude,
    represented by the shaded circle. The single line denotes the
    single boson, while the double line denotes the diboson field.}
  \label{fig:inteq}
\end{figure}

The central quantity in the three-body sector is the boson-diboson
amplitude. The derivation of the equation governing this object in
finite volume has been shown in detail in a previous
publication~\cite{Kreuzer:2009jp,phd} and is therefore only briefly
sketched in the following. A discussion of the infinite-volume case
can be found, e.g., in~\cite{Braaten:2004rn,Platter:2009gz}. Starting
point for both cases is the Lippmann-Schwinger equation for the
boson-diboson amplitude. In a finite volume, there is no continuous
scattering spectrum, but only discrete energy levels. The analytic
structure of the amplitude is given by a number of simple poles at the
energies of these discrete states, just like the analytic structure of
bound states in the infinite volume. In addition, the incoming and
outgoing quantum numbers separate. The residue at the pole is the
amplitude~$\mathcal{F}$. Comparing the residues of both sides of the
Lippmann-Schwinger equation at any of the discrete energy levels
yields the homogeneous equation depicted diagrammatically in
Figure~\ref{fig:inteq}.

In the center of mass frame, the momenta of the outgoing single boson
and diboson are~$\vec{p}$ and~$-\vec{p}$, respectively. The outgoing
single boson is taken to be on-shell, giving the
four-momentum~$(p^2/2, \vec{p})$. The four-momentum of the outgoing
diboson line is~$(E-p^2/2, -\vec{p})$, making the total kinetic
energy~$E$ a parameter of the equation. The equation depicted in
Fig.~\ref{fig:inteq} then serves as a consistency condition. Values of
the energy parameter~$E$ for which the equation has a solution are
identified as the energies of the discrete finite volume spectrum. 

Using the kinematics outlined above, the sum equation for the
boson-diboson amplitude~$\mathcal{F}$ reads
\begin{equation}\label{eq:sum}
  \mathcal{F}(\vec{p}) = \frac{8\pi}{L^3}
  \sum_{\vec{q}\in\frac{2\pi}{L}\zdrei}^\Lambda \mathcal{Z}(E; \vec{p},
  \vec{q}) \tau(E; q)\, \mathcal{F}(\vec{q}).
\end{equation}
The quantity $\mathcal{Z}(E; \vec{p}, \vec{q})$ contains the
interaction kernel of the sum equation, while $\tau(E; q)$ is
essentially the diboson propagator inside the loop.

For the diboson lines in the three-body equation depicted in
Fig.~\ref{fig:inteq}, the interacting diboson propagator has to be
used. This quantity is obtained by dressing the bare propagator, which
is a constant, with bosonic loops (cf.~\cite{Bedaque:1998km}). In the
sum equation~\eqref{eq:sum}, it appears via the
quantity~\cite{Kreuzer:2008bi,phd}
\begin{equation}\label{eq:diboson}
  \tau(E; q)=\bigg(\frac{1}{a}-\sqrt{\frac{3q^2}{4}-E} +
    \sum_{\genfrac{}{}{0pt}{}{\vec{\jmath}\in\zdrei}{\vec{\jmath}\neq\vec{0}}}
    \frac{1}{L|\vec{\jmath}|}\e^{-|\vec{\jmath}|L\sqrt{\frac{3q^2}{4}-E}}
    \bigg)^{-1},
\end{equation}
which is identical to the full propagator up to a prefactor. The
absolute value of the integer three-vector~$\vec{\jmath}$ is
denoted~$|\vec{\jmath}|$. In the limit $L\to\infty$, the propagator
reduces to the infinite volume expression.

The interaction kernel is given by the one-boson exchange term and the
boson-diboson contact interaction. Since the time direction is
considered to be of infinite extent, the integration over the loop
energy remains continuous and can be performed analytically by using
the residue theorem. The resulting interaction kernel depends only on
incoming and outgoing three-momenta as well as the total energy:
\begin{equation}
  \label{eq:z}
  \mathcal{Z}(E; \vec{p}, \vec{q}) = 
  \left[ \frac{1}{p^2 + \vec{p}\cdot\vec{q} + q^2 - E} 
    + \frac{H(\Lambda)}{\Lambda^2} \right]
\end{equation}

In the present calculation, only energies below the threshold for the
break-up into three individual bosons, i.e. $E < 0$, are considered.
This keeps the investigated states below any inelastic thresholds, in
particular the $Nd \rightarrow NNN$ threshold.

\subsection{Methodology}
The amplitude~$\mathcal{F}$ is strictly speaking only defined for the
discrete momenta inside the box. One ansatz for the solution of the
sum equation~\eqref{eq:sum} is therefore to use the quantized momenta
themselves as sampling points. In this approach, the value of the
amplitude at the sampling points is written as
$f_i=\mathcal{F}(\vec{p}_i)$, where the index $i$ labels all momenta
$p_i\in\frac{2\pi}{L^3}\zdrei$ with $|\vec{p}_i|<\Lambda$. This yields
the finite-dimensional eigenvalue equation
\begin{equation}
  \label{eq:grid}
  f_i = \sum_j  \left[ \frac{8\pi}{L^3} \mathcal{Z}(E; \vec{p}_i,\vec{q}_j)
  \tau(E; |\vec{q}_j|)\right] f_j 
\end{equation}
The range of box side lengths that allow for the use of this ansatz is
limited. The number of points to consider roughly scales as $(\Lambda
L)^3$, while the complexity of the necessary matrix diagonalization
scales like the number of points to the third power. We will therefore
only show results for $L \le a$ in this publication. Another drawback
of this approach is that it is cumbersome to disentangle the
contribution of different partial waves to the finite volume
spectrum. We will refer to this ansatz as the \emph{Grid approach}.

The momentum quantization inside a finite volume is tantamount to a
reduction of the spherical symmetry of the infinite volume to a
discrete symmetry, namely in our case the point symmetry of the
cube. In the language of group theory, the infinitely many irreducible
representations of the rotational group~SO(3) become reducible in
terms of the five irreducible representations of the cubic group~$O$.
Hence, a quantity~$\psi_s$ transforming according to the irreducible
representation~$s$ of~$O$ can be written in terms of the basis
functions of spherical symmetry, i.e. spherical harmonics~$Y_{\ell
  m}$, via
\begin{equation}
  \label{eq:kubic}
  \psi_s(\vec{p}) = \sum_{\ell, t} R_{\ell t}(p) K_{s\ell t}(\hat{p}),
  \text{ with } K_{s\ell t}(\hat{p}) = \sum_{m} C_{s\ell m}^{(t)}
      Y_{\ell m}(\hat{p}),
\end{equation}
where $\hat{p} = \vec{p} / p = (\hat{p}_1,\hat{p}_2,\hat{p}_3)$ is the
unit vector in $\vec{p}$-direction and~$R_{\ell t}$ is the radial
function. The index~$t$ is needed if the representation labeled
by~$\ell$ appears in the irreducible representation~$s$ more than
once. The coefficients~$C_{s\ell m}^{(t)}$ are real and normalized for
given values of $s$, $\ell$ and $t$ via $\sum_{m} \left[C_{s\ell
    m}^{(t)}\right]^2 = 1$. The linear combinations~$K_{s\ell t}$ of
spherical harmonics are called ``kubic harmonics''
(sic!)~\cite{vonderLage:1947zz}. The values of the
coefficients~$C_{s\ell m}^{(t)}$ are known for~$\ell$ as large
as~12~\cite{Altmann:1965zz} and are readily computed from group
theory~\cite{Bernard:2008ax}. In this work, we restrict ourselves to
amplitudes~$\mathcal{F}$ transforming under the trivial ($A_1$)
representation of the cubic group, because it contains the trivial
representation of spherical symmetry, namely $\ell=0$. This is because
the investigated Efimov states are S-wave states in the infinite
volume. In addition, $A_1$ receives contributions from $\ell =
4,6,8,\dots$ Since all $\ell$-values appear only once below $\ell=12$,
the index~$t$ is dropped in the following. The coefficients~$C_{A_1
  \ell m}$ for $\ell=0,4$ are summarized in Table~\ref{tab:clm}
together with the resulting kubic harmonics in Cartesian coordinates,
normalized to $4\pi$.

\begin{table}[t]
  \centering
  \addtolength{\tabcolsep}{.3cm}
  \renewcommand{\arraystretch}{1.2}
  \begin{tabular}{cc|c|c}
    \hline\hline
    $\ell$ & $m$ & $C_{A_1 \ell m}$ & $K_{A_1 \ell}(\hat{p})$ \\
    \hline\hline
    0 & 0 & 1 & 1 \\
    \hline
    4 & 0 & $\frac{\sqrt{21}}{6}$ & ~ \\
    4 & $\pm 1,2,3$ & 0 & $\frac{5\sqrt{21}}{4}\left(\hat{p}_1^4 + \hat{p}_2^4 + \hat{p}_3^4 -\frac{3}{5}\right)$ \\
    4 & $\pm 4$ & $\frac{\sqrt{30}}{12}$ & ~ \\
    \hline\hline
  \end{tabular}
  \caption{Coefficients $C_{A_1 \ell m}$ of the kubic 
    harmonics~\cite{Bernard:2008ax} as well as $K_{A_1 \ell}$ in 
    Cartesian coordinates~\cite{vonderLage:1947zz} for $\ell=0,4$.}
  \label{tab:clm}
\end{table}

The expansion of the angular dependence of the amplitude in spherical
harmonics allows for an assessment of the impact of higher partial
waves or, in other words, of the loss of spherical symmetry in the
finite volume. Moreover, it allows for an analytic calculation of the
angular dependence, leaving only the radial part for a numerical
treatment. This significantly reduces the complexity of the numerical
problem at hand compared to the Grid approach. We therefore determine
the $\hat{p}$-dependence of the right hand side of
Eq.~\eqref{eq:sum}. The only quantity where $\vec{p}$ is present is
the one-boson exchange part of the interaction
kernel~$\mathcal{Z}(E;\vec{p},\vec{q})$, given by the first term in
Eq.~\eqref{eq:z}. Consider therefore the quantity
\begin{equation}
  \label{eq:ipqdef}
  I_p^{(\ell)}(\vec{q}) = \int_{S^2} \frac{\dd^2\hat{p}}{4\pi}\,
    K_{A_1 \ell}(\hat{p})\left[\sum_{C\in O}
      \left(\frac{1}{\alpha + \vec{p}\cdot C\vec{q}} + 
        \frac{1}{\alpha - \vec{p}\cdot C\vec{q}}\right)\right],
\end{equation}
where we have set $\alpha=p^2+q^2-E$ for simplicity. The summation of
all elements of the cubic group~$O$ makes the expression in square
brackets invariant under any cubic rotation of~$\vec{p}$
and~$\vec{q}$. It therefore transforms according to the trivial
representation, denoted $A_1$, of the cubic group. The inclusion of
the second term makes it moreover invariant under additional parity
transformations. The expansion coefficients of this quantity in kubic
harmonics are given by~$I_p^{(\ell)}(\vec{q})$. The analytic
evaluation of the angular integral is shown in
Appendix~\ref{app:ipq}. The result is
\begin{equation}
  \label{eq:ipq}
  I_p^{(\ell)}(\vec{q}) = \frac{48}{pq}Q_\ell\left(\frac{\alpha}{pq}\right) 
  K_{A_1 \ell}\left(\hat{q}\right),
\end{equation}
where $Q_\ell$ is the $\ell$th Legendre function of the second kind as
defined in~\cite{Gradshteyn:7ed}. Special care has to be taken of the
limiting cases $p,q\to 0$. The result
$I_0(\vec{q})=I_p(\vec{0})=\frac{48}{\alpha}\delta_{\ell 0}$ coincides
with the zero-momentum limit of~\eqref{eq:ipq}.

We will now use this result to simplify -- at least from a numerical
viewpoint -- the sum equation~\eqref{eq:sum}. The summation over the
discrete set of three-vectors can be rewritten in terms of a summation
over the elements of the cubic group as follows:
\begin{equation}
  \label{eq:rewritesum}
  \sum_{\vec{q}\in\frac{2\pi}{L}\zdrei}f(\vec{q}) = 
  \sum_{\vec{q}\in\frac{2\pi}{L}\langle\zdrei\rangle}\frac{\mul(\vec{q})}{48}
  \sum_{C \in O}\left(f(C\vec{q})+f(-C\vec{q})\right),
\end{equation}
where $\langle\zdrei\rangle = \{\vec{q}\in\zdrei : q_1 \ge q_2 \ge q_3
\ge 0\}$. All other vectors in~$\zdrei$ can be generated by applying
cubic rotations and the parity operator to the vectors in
$\langle\zdrei\rangle$, i.e., 
\begin{equation}
  \label{eq:allvectors}
  \bigcup_{C \in O} (C + CP) \, \langle\zdrei\rangle = \zdrei.
\end{equation}
However, the summation over all cubic rotations
in~\eqref{eq:rewritesum} might introduce double
counting. Specifically, if the original vector contains identical
entries or zeroes, several cubic rotations have the same image
vector. This is accounted for by the factor ${\mul(\vec{q})}/{48}$,
where $\mul(\vec{q})$ is the number of vectors that contain the same
entries as $\vec{q}$ up to ordering and signs. The identity
Eq.~\eqref{eq:rewritesum} was also used to reduce the dimensionality
in the numerical implementation of the Grid
approach. Applying~\eqref{eq:rewritesum} to Eq.~\eqref{eq:sum} yields
\begin{equation}
  \label{eq:angsum}
  \begin{split}
    \mathcal{F}(\vec{p}) &= \sum_\ell^{(A_1)}R_\ell(p)K_{A_1 \ell}(\hat{p})
    = \frac{8\pi}{L^3} \sum_{\vec{q}\in\frac{2\pi}{L}\langle\zdrei\rangle}
    \frac{\mul(\vec{q})}{48} \tau(E;q) \mathcal{F}(\vec{q}) \sum_{C \in O}
    \left(\mathcal{Z}(E;\vec{p},C\vec{q})+\mathcal{Z}(E;\vec{p},-C\vec{q})\right)\\
    &= \frac{8\pi}{L^3} \sum_{\vec{q}\in\frac{2\pi}{L}\langle\zdrei\rangle}
    \left(\frac{1}{pq}\sum_\ell^{(A_1)}
    Q_\ell\left(\frac{p^2+q^2-E}{pq}\right)K_{A_1 \ell}(\hat{p})K_{A_1 \ell}(\hat{q}) + 
    \frac{H(\Lambda)}{\Lambda^2}K_{A_1 0}(\hat{p})\right) \\
  &\hspace{3.5cm}\times\mul(\vec{q})\tau(E;q) \mathcal{F}(\vec{q}),
  \end{split}
\end{equation}
where the $\ell$-sums run over the partial waves contained in the
$A_1$-representation, denoted by the upper summation limit~$(A_1)$, and
$\mathcal{F}(C\vec{q})=\mathcal{F}(\vec{q})\; \forall\, C\in O$ was
used in the first line.

Comparing the coefficients of $K_{A_1 \ell}(\hat{p})$ in
Eq.~\eqref{eq:angsum} yields a set of coupled equations for the radial
functions~$R_\ell$. Let $R_{\ell i} = R_\ell(p_i)$, where the~$p_i$
are the possible absolute values of vectors in~$\frac{2\pi}{L}\zdrei$
smaller than the cutoff~$\Lambda$. The resulting expression has the
form of an eigenvalue problem
\begin{equation}
  \label{eq:eigen}
  R_{\ell i} = M_{\ell i;\lp j}(L; E)\, R_{\lp j},
\end{equation}
where the matrix~$M$ is given by
\begin{equation}
  \label{eq:kernel}
  \begin{split}
    M_{\ell i;\lp j}(L; E) =& \frac{8\pi}{L^3} 
    \sum_{\stackrel{\vec{q}\in\frac{2\pi}{L}\langle\zdrei\rangle}{|\vec{q}|=q_j}}^{\Lambda}
    \left[\frac{1}{p_i q_j}Q_\ell\left(\frac{p_i^2+q_j^2-E}{p_i q_j}\right)
      K_{A_1 \ell}(\hat{q}) + \frac{H(\Lambda)}{\Lambda^2}\delta_{\ell 0}\right]\\
    &\hspace{3.5cm}\times
    \mul(\vec{q})\tau\left(E;q_j\right) K_{A_1 \lp}(\hat{q}).
  \end{split}
\end{equation}
We use a root finding algorithm to determine values of the energy
parameter~$E$ such that the eigensystem~\eqref{eq:eigen} contains an
eigenvalue~1, yielding the discrete energy levels inside the finite
volume.

For the discussion of the perturbative approaches, it will be
convenient to write the matrix equation~\eqref{eq:eigen} in block
matrix form in angular momentum space, viz.
\begin{equation}
  \label{eq:block}
  \renewcommand{\arraystretch}{1.75}
  \addtolength{\arraycolsep}{\arraycolsep}
  \left(\begin{array}{c} R_0 \\ \hline R_4 \\ \hline \vdots \end{array}\right) =
  \left(\begin{array}{c|c|c} M_{00} & M_{04} & \dots \\ \hline
      M_{40} & M_{44} & \dots \\ \hline 
      \vdots & \vdots & \ddots \end{array}\right)\,
  \left(\begin{array}{c} R_0 \\ \hline R_4 \\ \hline \vdots \end{array}\right),
\end{equation}
with $(R_\ell)_i = R_{\ell i}$ and $(M_{\ell\lp})_{ij} = M_{\ell i;
  \lp j}$. The dependence of the block matrices on the box side length
and the energy parameter is suppressed in this notation. The
off-diagonal block matrices in~\eqref{eq:block} are responsible for
the mixing of different partial waves.

We will refer to the method described above as the \emph{Sum
  approach}.  The conceptual advantage of this approach is the
possibility to disentangle the contributions of the different partial
waves to the energy shift in the finite volume. On the numerical side,
the dimensionality of the eigenvalue problem is greatly reduced in
comparison to the Grid approach: if there are several three-momenta
with identical absolute value, they all increase the dimensionality of
the Sum approach only by one. The differences in dimensionality and
runtime will be discussed at the end of Section~\ref{results}.

In earlier works~\cite{Kreuzer:2008bi,Kreuzer:2009jp,phd} the loop sum
was rewritten into a sum of integrals via the Poisson equation. This
explicitly recovers the infinite volume form of the amplitude. This is
appropriate for bound states, where this approach was used. It is,
however, not appropriate when extending the formalism into the energy
region of elastic boson-diboson scattering. Here, the infinite and
finite volume amplitudes have fundamentally different analytic
structures, namely a scattering continuum and a series of poles
corresponding to a discrete spectrum, respectively. This energy region
is accessible with the new Sum approach. We will refer to the older
approach as the \emph{Poisson approach}.

\subsection{Perturbative approaches}
In the infinite volume, bound states and low-lying scattering states
are predominantly S-wave states.  In the finite volume, there are
admixtures from higher partial waves stemming from the breakdown of
the spherical symmetry, as discussed above. For reasonably large
volumes, these admixtures are expected to be small compared to the
S-wave only part.

To see that this is indeed the case, consider the argument of the
Legendre function~$Q_\ell$ in the definition of the matrix
elements~\eqref{eq:kernel}. The argument $(p_i^2+q_j^2-E)/(p_iq_j)$ is
guaranteed to be larger than~1 (remember that $E<0$ here) and will be
large if the ratio $p_i/q_j$ is either very large or very small. For
large arguments $z \gg 1$, the Legendre functions $Q_\ell(z)$ scale as
$z^{-\ell-1}$~(see, e.g.,\cite{Gradshteyn:7ed}). Therefore,
\begin{equation}
  \label{eq:ldepql}
  \frac{Q_4(z)}{Q_0(z)} \sim z^{-4}\text{, if } z \gg 1,
\end{equation}
which shows the suppression of the G-wave contributions. The momenta
$p_i,q_j$ are multiples of the low-momentum scale $2\pi/L$. Consider,
for example, the pair of momenta with the largest possible ratio,
namely $\Lambda$ and $\frac{2\pi}{L}$. With these choices, and the
condition $\Lambda L \gg 1$ from renormalizability, the argument of
the Legendre function becomes
$\frac{L}{2\pi}(\Lambda+(E_3/\Lambda))$. Using the expansion of the
Legendre functions for large arguments, we can estimate the scaling of
the G-wave correction by
\begin{equation}
  \label{eq:ldepscale}
  \frac{Q_4}{Q_0}\sim\left(\frac{2\pi}{\Lambda L}\right)^4.
\end{equation}
The scaling of other matrix elements with the box side length is not
as drastic but still governed by the strong suppression evidenced in
Eq.~\eqref{eq:ldepql}.

A second suppression mechanism comes from the kubic harmonics in the
definition of the matrix elements~\eqref{eq:kernel}. While the kubic
harmonic $K_{A_1 0}$ is just a constant, the corresponding functions
for $\ell\neq 0$ show sign oscillations
(cf. Table~\ref{tab:clm}). These lead to cancellations among the terms
from vectors with the same absolute value. These corrections are
themselves independent of the size of the volume as they reflect the
breakdown of the spherical symmetry. However, there are more
cancellations among integer vectors of high absolute values, where the
number of vectors with identical absolute value is large. These
vectors are reached for $\Lambda L \gg 1$, leading to more
cancellations within the summation over the possible momenta.

Overall, the two suppression mechanisms in the definition of the
matrix elements introduce the generic hierarchy
\begin{equation}
\label{eq:hierarchy}
M_{00} \gg M_{04} \gg M_{40} \gg M_{44} \gg M_{60} \gg \dots
\end{equation}
In the following, we discuss two different approaches to take
advantage of this ordering. The first one can be described as a
partially resummed calculation, while the second approach is developed
as a strict perturbative expansion. In the following discussion, we
will use the notation introduced in Eq.~\eqref{eq:block}.

For the \emph{Partial Resummation approach}, consider the second row
of Eq.~\eqref{eq:block},
\begin{equation}
  \label{eq:r4}
  R_4 = M_{40} R_0 + M_{44} R_4 + \dots
\end{equation}
Inserting this expression for~$R_4$ into the equation for~$R_0$ given
by the first row of Eq.~\eqref{eq:block} yields the
term~$M_{04}M_{44}R_4$ containing an additional factor of $K_{A_1
  4}(\vec{q})$ compared to~$M_{04}M_{40}R_0$. This term is therefore
discarded as parametrically small in the partial resummation
approach. More generally, the partial wave amplitudes~$R_\ell$ for
$\ell\neq 0$ are approximated by $R_{\ell\neq 0} = M_{\ell 0}R_0$, or
equivalently by setting all $M_{\ell\neq 0,\lp\neq 0}$ to zero:
\begin{equation}
  \label{eq:resum}
  \renewcommand{\arraystretch}{1.75}
  \addtolength{\arraycolsep}{\arraycolsep}
  \left(\begin{array}{c} R_0 \\ \hline R_4 \\ \hline \vdots \end{array}\right) \simeq
  \left(\begin{array}{c|c|c} M_{00} & M_{04} & \dots \\ \hline
      M_{40} & 0 & 0 \\ \hline 
      \vdots & 0 & 0 \end{array}\right)\,
  \left(\begin{array}{c} R_0 \\ \hline R_4 \\ \hline \vdots \end{array}\right)
\Longrightarrow 
R_0 \simeq \left( M_{00} + \sum_{\ell\neq 0} M_{0\ell} M_{\ell 0}\right)R_0,
\end{equation}
where the sum over $\ell$ is truncated at $\ell=0$ or $\ell=4$ in the
calculations presented in this paper. This is possible due to the
general hierarchy~\eqref{eq:hierarchy}. Using this equation, the energy
of the state can be obtained by the eigenvalue method described above.

For the \emph{Strictly Perturbative approach}, we first establish how
to obtain the perturbed energies in the eigenvalue method. Consider
the general eigenvalue equation
\begin{equation}
  \label{eq:peval}
  T(E) = K(E) \, T(E)
\end{equation}
for an amplitude $T$ and a matrix $K$, where both quantities depend on
a parameter~$E$ that is tuned such that $T$ is an eigenvector of $K$
with eigenvalue 1. If the matrix~$K$ contains a small perturbation
$K(E)=K_0(E)+\varepsilon K_1(E)$ with $\varepsilon \ll 1$, both~$T(E)$
and~$E$ can be expanded in powers of~$\varepsilon$, retaining only the
linear term:
\begin{equation}
  \label{eq:pamperg}
  \begin{split}
    T(E) &= T_0(E) + \varepsilon T_1(E) + \mathcal{O}(\varepsilon^2), \\
    E &= E_0 + \varepsilon E_1 + \mathcal{O}(\varepsilon^2),
  \end{split}
\end{equation}
such that
\begin{equation}
  \label{eq:plo}
    T_0(E_0) = K_0(E_0) \, T_0(E_0)
\end{equation}
holds. Furthermore, the $E$-dependence of~$K_0$ and~$T_0$ can be
linearized around~$E_0$ as
\begin{equation}
  \label{eq:pexp}
  \begin{split}
    K_0(E) &= K_0(E_0) + \varepsilon E_1 K_0^\prime(E_0)
                   + \mathcal{O}(\varepsilon^2) \\
    T_0(E) &= T_0(E_0) + \varepsilon E_1 T_0^\prime(E_0)
                   + \mathcal{O}(\varepsilon^2),
  \end{split}
\end{equation}
where the prime indicates differentiation with respect to the energy
parameter. Inserting the expansions~\eqref{eq:pamperg}
and~\eqref{eq:pexp} into~\eqref{eq:peval} and comparing orders
of~$\varepsilon$ yields to order~$\varepsilon^0$ just
Eq.~\eqref{eq:plo}. To order~$\varepsilon^1$, we obtain
\begin{equation}
  \label{eq:pnlo}
  E_1\,T_0^\prime + T_1 = E_1 \, K_0^\prime T_0 
    + K_1 T_0 + K_0\left[E_1\,T_0^\prime + T_1\right],
\end{equation}
where all quantities are evaluated at the leading order
energy~$E_0$. Multiplying by~$T_0^\dagger(E_0)$ from the left hand
side and using the daggered version of Eq.~\eqref{eq:plo} yields the
desired equation for the energy shift in lowest order in~$\varepsilon$
\begin{equation}
  \label{eq:pdeltae}
  E_1 = - \frac{T_0^\dagger K_1 T_0}{T_0^\dagger K_0^\prime T_0}.
\end{equation}
Note that the leading order amplitude~$T_0$ appears only at the
leading order energy.

In the present framework, the leading order matrix~$K_0$ is given by
the block matrix~$M_{00}$ in Eq.~\eqref{eq:block}, while the leading
order amplitude is accordingly the amplitude~$R_0$ determined such
that
\begin{equation}
  \label{eq:swave}
  R_0(E_0) = M_{00}(E_0) R_0(E_0)
\end{equation}
holds, cf. Eq.~\eqref{eq:plo}. Since the dependence of~$M_{00}$ on the
energy parameter is known from Eq.~\eqref{eq:kernel}, the
differentiation in the denominator of~\eqref{eq:pdeltae} can be
carried out analytically. As can be read off from Eq.~\eqref{eq:resum}
in the discussion of the partial resummation approach, the
perturbation is given by
\begin{equation}
  \label{eq:ppert}
  K_1 =  \sum_{\ell\neq 0} M_{0\ell} M_{\ell 0},
\end{equation}
where the sum over~$\ell$ will again be truncated at $\ell=4$ in our
calculations.

In earlier publications~\cite{Kreuzer:2008bi,Kreuzer:2009jp,phd}, we
showed how the computational effort can be reduced by expanding the
matrix~$M_{00}(E)$ around the infinite volume energy. Results from
these calculations agree with those obtained using the full energy
dependent matrix as long as the shift from the infinite volume energy
is smaller than 20\%. However, since these shifts are in general much
larger for volume sizes typical for present day lattice calculations,
this earlier approach is not applicable in the present work and
therefore not followed further.


\section{Results and discussion}
\label{results}

Using the formalism laid out in the previous section, we now present
the negative energy spectrum in finite cubic volumes of varying side
lengths. We provide a detailed discussion of the results with emphasis
on the applicability of the perturbative approaches. For convenience,
the dependence of the energies on the boson mass~$m$ is reinstated in
the following.


In this publication, we focus on systems with a positive two-body
scattering length~$a$. In such systems, a physical diboson state with
the energy $E_D = -1 / (ma^2)$ exists in the infinite volume. This
energy constitutes the threshold for the break-up of a triboson into a
diboson and a single boson. The volume dependence of the diboson in a
finite cubic box is derived in~\cite{Luscher:1990ux,Beane:2003da}.
The threshold for the break-up into three single bosons is given by
$E=0$.

We investigate four states with different energies in the infinite
volume:
\begin{itemize}
\item {\bf State A: } $E_3^\infty = -10 / (ma^2)$
\item {\bf State B: } $E_3^\infty = -5 / (ma^2)$
\item {\bf State C: } $E_3^\infty = -3 / (ma^2)$
\item {\bf State D: } $E_3^\infty = -1.5 / (ma^2)$
\end{itemize}
In these ``universal units'', the energy of the three-body state is
given in units of the two-body bound state, which is physical for $a >
0$ and virtual for $a < 0$.

For each of these states, the finite volume spectrum has been
calculated for several values of the box side length~$L$. In
applications of the formalism to Lattice QCD, the volume under
consideration will be smaller than the unnaturally large scattering
length. 
Therefore, we concentrate on the region $L \le a$ and show
results for various box side lengths between $L=a$ and
$L=0.25a$. 

In order to provide an understanding of the universal units employed
in this publication, we take a look at the nucleonic sector. There,
two spin-isospin channels with associated scattering lengths
$a_{^1S_0}=-23.7$~fm and $a_{^3S_1}=5.4$~fm are present. Using a
``middle ground'' of 15~fm for the unnaturally large scattering
length, we see that the investigated volume range contains typical
volume sizes of present day Lattice QCD calculations, which are about
2.5~fm to 4~fm. The triton energy in units of the deuteron binding
energy is $E_T=-3.8 E_D$, placing this state roughly between the
states~B and~C. Note, however, that all results presented in this
paper are for bosonic systems.

In order to explicitly verify that our results are renormalized, we
use different cutoffs~$\Lambda$. For all cutoffs considered, the
three-body force parameterized by~$H(\Lambda)$ has been fixed such
that the same infinite volume energy is obtained. If our results are
indeed renormalized, the results for the different cutoffs should
agree with each other up to an uncertainty of the order~$1/(\Lambda
a)$ stemming from the finiteness of the cutoff. In lattice
simulations, the scale for the cutoff is set by the edge of the first
Brillouin zone given by $\pi/b$, where $b$ is the lattice spacing. The
cutoffs employed in this calculation, $\Lambda a = 314$ and $\Lambda a
=225$, correspond to typical present day lattice spacings of
$b=0.1$~fm and $b=0.15$~fm, respectively.

We investigated five different variants. The Grid approach from
Eq.~\eqref{eq:grid} is the exact result. All other runs employ the Sum
approach. The leading order of our perturbative approaches is
obtained by only taking S-wave contributions into account
(cf. Eq.~\eqref{eq:swave}). The three remaining runs include the
G-wave ($\ell=4$) using the full equation~\eqref{eq:eigen} as well as
the two perturbative approaches (cf. Eqs.~\eqref{eq:resum},
\eqref{eq:pdeltae}) described above.

\begin{figure*}[t]
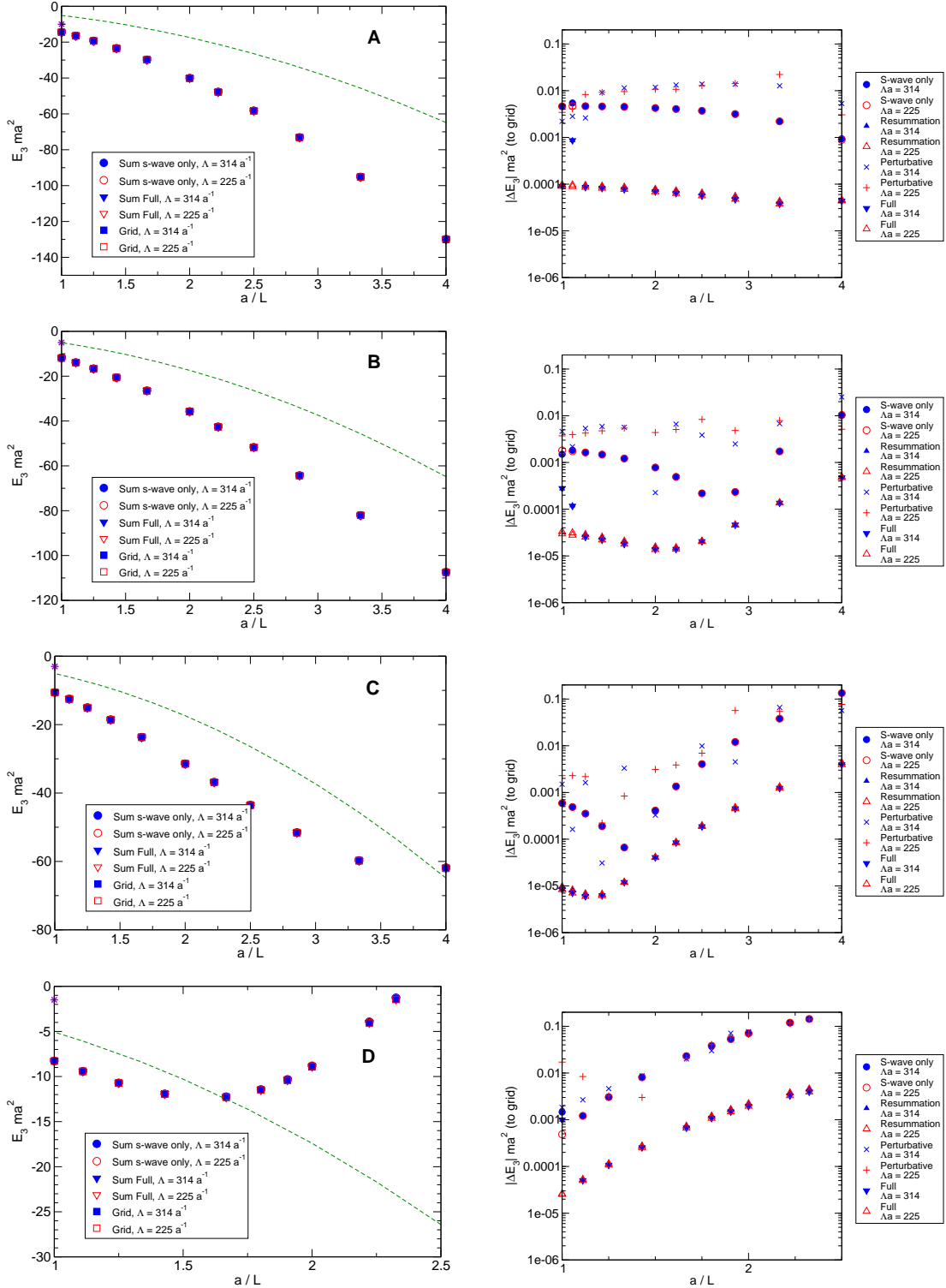

  \centering
  \includegraphics*[width=.42\linewidth]{zehn.eps}
  \hspace{.5cm}
  \includegraphics*[width=.42\linewidth]{diff_zehn.eps}\\[.3cm]
  \includegraphics*[width=.42\linewidth]{fuenf.eps}
  \hspace{.5cm}
  \includegraphics*[width=.42\linewidth]{diff_fuenf.eps}\\[.3cm]
  \includegraphics*[width=.42\linewidth]{drei.eps}
  \hspace{.5cm}
  \includegraphics*[width=.42\linewidth]{diff_drei.eps}\\[.3cm]
  \includegraphics*[width=.42\linewidth]{andert.eps}
  \hspace{.5cm}
  \includegraphics*[width=.42\linewidth]{diff_andert.eps}
  \caption{\emph{Left:} Variation of the three-body energy $E_3$ with
    the box side length $L$ for two cutoffs. Plotted are the Sum
    approach with S-wave only and including the G-wave as well as the
    Grid approach. The infinite volume energy is marked as a star on
    the left axis. The dashed line is the two-body energy.
    \emph{Right:} Difference between the Grid approach and the Sum
    approach results (S-wave only and including the G-wave via full
    calculation, partial resummation and strict perturbation
    theory). Note the different volume range in the last row.}
  \label{fig:voldep}
\end{figure*}

The volume dependence of the investigated states is depicted in the
left panels of Fig.~\ref{fig:voldep} for the Grid approach as well as
the S-wave and full G-wave calculation in the Sum approach. The
infinite volume energy of the state is shown as a star on the left
axis.  The curves for different cutoffs agree with each other to
within better than 1\%, indicating proper renormalization of the
results. There is no visible difference between the three depicted
variants for all investigated volumes.

The curves show the qualitative behavior known from our previous work
for box side lengths down to
$L=0.5a$~\cite{Kreuzer:2008bi,Kreuzer:2009jp,phd}. When decreasing the
volume size from infinity, the state remains unaffected at
first. Going below a certain box size starts to affect the state
strongly, which is in line with the expected exponential behavior of
finite volume corrections for bound states. The box size below which
the state shows strong deviations from its infinite volume energy is
tied to the spatial extent of the state in the infinite volume, which
can be estimated as $(-mE_3^\infty)^{-1/2}$.  The size of the state is
also correlated with the length scale in the exponential behavior of
the finite volume corrections.

\begin{table}[t]
  \centering
  \addtolength{\tabcolsep}{.3cm}
  \begin{tabular}{c|c|c||c|c|c}
    \hline\hline
    ~ & ~ & ~ & $\Lambda a = 225$ & $\Lambda a = 314$ & ~ \\
    State & $E_3^\infty\,ma^2$ & $(-mE_3^\infty)^{-1/2}$ &
    $E_3(L=a)\,ma^2$ & $E_3(L=a)\,ma^2$ & $\delta_\text{rel}$ \\
    \hline\hline
    A & -10 & 0.32~$a$ & -14.49 & -14.49 & 44.9\% \\
    B & -5 & 0.45~$a$ & -11.87 & -11.87 & 137.4\% \\
    C & -3 & 0.58~$a$ & -10.63 & -10.63 & 254.3\% \\
    D & -1.5 & 0.82~$a$ & -8.28 & -8.28 & 452.0\% \\
    \hline\hline
  \end{tabular}
  \caption{Comparison of infinite volume energies and finite volume 
    energies at $L=a$ as calculated by the Grid approach. Shown are 
    results for the two different cutoffs along with the relative 
    deviation from the infinite volume 
    energy~$\delta_\text{rel}=(E_3(L=a)/E_3^\infty)-1$. Also shown is 
    the size estimate $(-mE_3^\infty)^{-1/2}$.}
  \label{tab:leqa}
\end{table}
To see that this is indeed the case, we summarize the finite volume
shifts at $L=a$ calculated using the Grid approach for all four states
and the two investigated cutoffs Table~\ref{tab:leqa}. The results for
the two cutoffs agree with each other, again showing that the finite
volume results are renormalized.  Comparing the shifts, we notice that
state~A is shifted by 45\%, while the more shallow and, therefore,
larger state~B is already shifted by 137\%.  For all states except the
deepest state~A the relative deviation from the infinite volume energy
exceeds 100\%. It is even larger than 400\% for the shallowest
state~D. The shift at $L=a$ is strongly dependent on the size of the
state in the infinite volume, and therefore on the infinite volume
energy. Thus, a given finite volume affects a shallow state with a
large spatial extent more than smaller, more deeply bound state, as
expected.

The volume dependence of the three-body energy as shown in the left
panels of Fig.~\ref{fig:voldep} seems identical for the three plotted
approaches. However, there are small differences between the different
runs. We use the exact result from the Grid approach as a baseline and
also present the differences $\Delta E_3(L) = E_3^\text{(Sum)}(L) -
E_3^\text{(Grid)}(L)$ in the right panels of
Fig.~\ref{fig:voldep}. Here, $E_3^\text{(Sum)}$ can be any of the four
runs employing the Sum approach. Namely, these are the S-wave only
calculation, the full calculation including the G-wave and the two
perturbative approaches.

The differences to the respective baseline appear cutoff
independent. Some variation is visible in the Strictly Perturbative
treatment, but it is within the uncertainty from the finiteness of the
cutoff which is of the order $1/(\Lambda a)$.

The full results with and without the G-wave as well as the partially
resummed results are always larger than the Grid results.  The S-wave
only result in general deviates from the Grid result by more than
0.001\ergunit. The inclusion of the G-wave consistently reduces the
difference to the full Grid result by about two orders of
magnitude. In other words, the inclusion of higher partial waves
lowers the energy of the investigated state.

For the very deeply bound state~A, the difference between the
respective Sum approach runs and the Grid approach does only weakly
depend on the volume. For volumes with side lengths larger than
$L=a/2$, state~B shows the same qualitative behavior. Below $L=a/3$,
the difference between the Sum approach and the Grid approach becomes
larger, but is still very small compared to the energy of the state
itself. This can be understood as higher partial waves with $\ell\ge
6$ having a larger and larger contribution. In addition, the absolute
size of the G-wave contribution strongly increases, as expected, when
going to smaller volumes.

Turning to state~C, we observe a behavior similar to that of
state~B. The strong increase in the spread between the Sum approach
runs and the Grid approach as well as in the size of the G-wave
contribution starts at a larger volume, namely around $L=0.6a$. This
reflects the larger spatial extent of the more shallowly bound state~C
compared to state~B. The differences of the Sum approach to the Grid
results grow by two orders of magnitude within the investigated volume
range. For the full result, for example, the difference increases from
$10^{-5}$\ergunit{} at $L=a$ to 0.004\ergunit{} at $L=a/4$. For the
very shallow state~D, we observe a strong increase of the differences
when going towards smaller volumes. They also grow by two orders of
magnitude, albeit over a much smaller volume range. The difference of
the full result at $L=a$ is $2\times 10^{-5}$\ergunit, about twice as
large as for state C. It has grown by two orders of magnitude to
0.004\ergunit{} already at $L=0.43a$.

The size of the higher partial wave contributions is comparable to the
uncertainty due to the finiteness of the cutoff, which is of order
$1/(\Lambda a)$. This should change when including higher orders of
the EFT. The next-to-leading order corrections are governed by the
effective range $r_e$ and are of the order $r_e / a$ and $kr_e$, where
$k$ is a typical momentum. Their inclusion will be part of the
extension of this framework to the three-nucleon system.

Overall, the G-wave and higher corrections are indeed small compared
to the S-wave energies, justifying their perturbative treatment. In
particular, the partial resummation approach captures the G-wave
corrections almost completely. More than 99\% of the shift from the
S-wave result to the full result are accounted for using this
approach.

For sufficiently large box sizes, states~C and~D show a different
volume dependence than the more deeply bound states (bottom left
panels of Fig.~\ref{fig:voldep}). For both states, we find energies
above the two-body energy depicted by the dashed line. For state~C,
this is the case for the smallest investigated volume $L=a/4$
only. The data points for $L=0.3a$ suggest a flattening out of the
volume dependence, allowing the two-body energy to catch up when
further decreasing the box size. This is more clearly seen in the
volume dependence of state~D. At $L=0.6a$, the two- and three-body
energy in the finite volume are equal. For smaller volumes, the
three-body state is found above the two-body state. Decreasing the
volume further increases the energy of the state, until it reaches the
three-body breakup threshold $E=0$ for a volume with side length
$L=0.43a$. We also observed this behavior in the framework developed
earlier~\cite{Kreuzer:2009jp,phd}.

Our interpretation is that, as the two-body energy reaches the
three-body energy, the state behaves like a boson-diboson scattering
state when further shrinking the volume. Indeed, the data points above
the threshold are compatible with a power-law behavior that is known
to describe two-body scattering. Note, however, that we are not in the
L{\"u}scher limit $L \gg a$ and therefore do not expect a leading
$1/L^3$ behavior~\cite{Luscher:1990ux}. The presence of three-body
effects in the volume dependence of such a state will be the subject
of further study. It was recently shown that, in this case, a
topological phase is to be expected~\cite{Bour:2011ef}.

The volume size at which this crossing happens is again tied to the
infinite volume energy. State D has a smaller infinite volume binding
energy, and therefore a larger spatial extent. Thus, the box which is
not able to hold the complete three-body bound state is larger than
for the smaller state~C.

Turning back to the differences shown in the right panels of
Fig.~\ref{fig:voldep}, we note that the volume dependence of the
higher partial wave contributions does not show any distinctive
features at the crossover box sizes.

\begin{figure*}[t]
  \centering
  \includegraphics*[width=.66\linewidth]{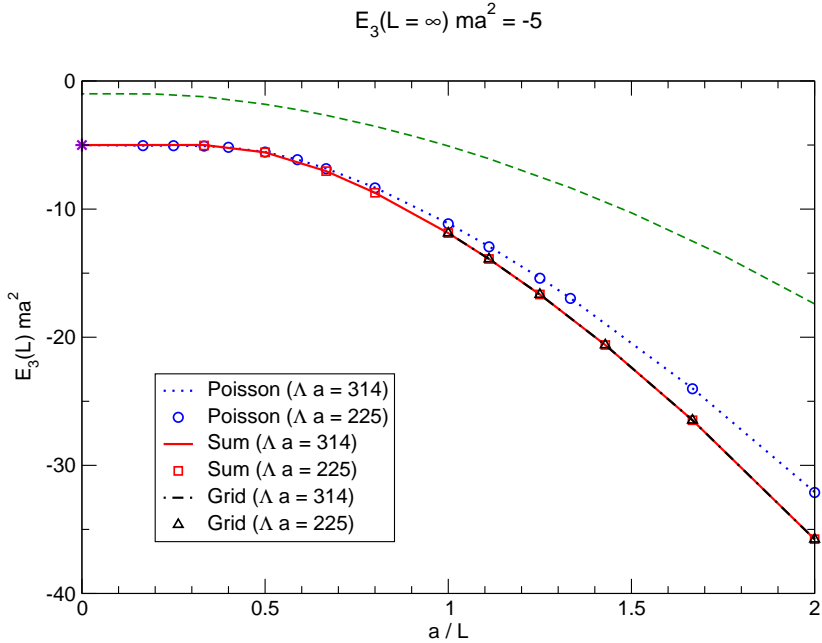}
  \caption{Volume dependence of state~B for two cutoffs, $\Lambda a =
    314$ (lines) and $\Lambda a = 225$ (symbols). Shown are the exact
    result from the Grid approach as well as the S-wave only results
    from the Sum approach and the Poisson approach. The infinite
    volume energy $E_3^\infty\, ma^2 = -5$ is marked as a star on the
    left axis. The dashed line is the two-body energy.}
  \label{fig:poisson}
\end{figure*}

In the following, we compare the Poisson approach from
Refs.~\cite{Kreuzer:2009jp,Kreuzer:2010ti,phd} to the approaches
developed in the present paper. The values for the Poisson and Sum
approach in Fig.~\ref{fig:poisson} have been obtained by truncating
the partial wave expansion of the angular dependence at $\ell=0$. We
observe a visible spread between the Sum approach and the Grid
approach on the one hand and the Poisson approach on the other. In the
Poisson approach, the sum equation~\eqref{eq:sum} is rewritten as
\begin{equation}
  \label{eq:poisson}
  \begin{split}
    \mathcal{F}(\vec{p}) &\stackrel{\phantom{(Poisson)}}{=}
      \frac{8\pi}{L^3} \sum_{\vec{q}\in\frac{2\pi}{L}\zdrei}^\Lambda
      \mathcal{Z}(E; \vec{p}, \vec{q}) \tau(E; q)\,
      \mathcal{F}(\vec{q}) \\
      & \stackrel{\text{(Poisson)}}{=} \frac{1}{\pi^2}
      \sum_{\vec{n}\in\zdrei}\int_0^\Lambda \dd q\,
      \left[\int\dd^2\hat{q} \, \e^{iL\vec{n}\cdot\vec{q}} 
        \mathcal{Z}(E; \vec{p}, \vec{q})\right]
      \tau(E; q)\, \mathcal{F}(\vec{q}).
    \end{split}
\end{equation}
The additional angular dependence introduced by the Fourier
transformation leads to a recoupling of the partial waves in this
approach. We note again that the S-wave component of the Sum approach
almost completely captures the full result, as discussed before. In
the Poisson approach, in contrast, the higher partial wave
contributions are sizable.

This is encouraging in view of the extension of this framework to the
three-nucleon system, which will be the subject of a future
publication. In the old framework, we compared the volume dependence
of the triton~\cite{Kreuzer:2010ti,phd} to results of a Lattice Chiral
EFT calculation~\cite{Epelbaum:2009zsa}. The decline of the energy
observed in the latter was much stronger. The results described above
indicate that this discrepancy will be at least diminished when
extending the present framework to the three-nucleon system.

\begin{figure}[t]
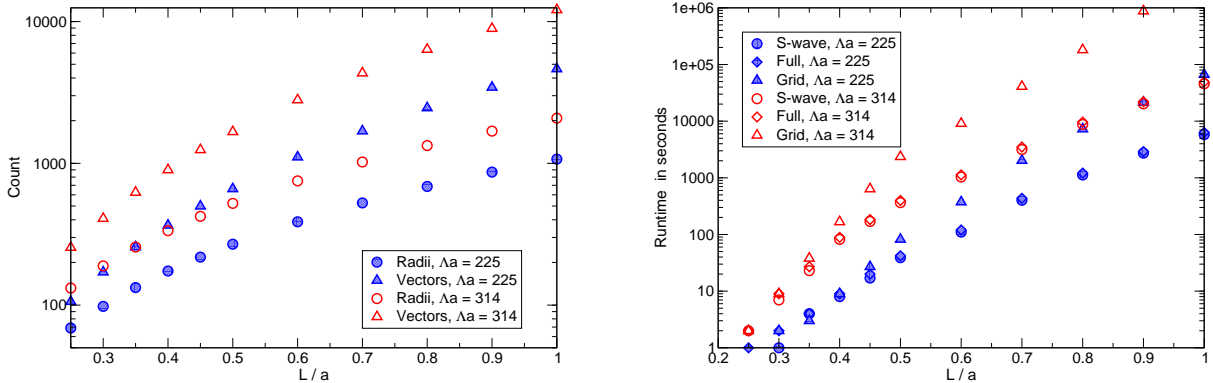

  \centering
  \includegraphics[width=.45\linewidth]{vectors.eps}
  \hspace{.95cm}
  \includegraphics*[width=.45\linewidth]{runtime.eps}
  \caption{Left: Number of vectors
    $\vec{q}\in\frac{2\pi}{L}\langle\zdrei\rangle$ with $|\vec{q}| <
    \Lambda$ and number of distinct absolute values for the two
    investigated cutoffs. Right: Length dependence of the runtime for
    three runs and two cutoffs.}
  \label{fig:count}
\end{figure}
We close this section with a discussion of the numerical scaling of
the presented approaches. The most costly operation in this framework
is finding eigenvalues, which is equivalent to performing a matrix
diagonalization. This is in general an $\mathcal{O}(N^3)$ process for
real non-symmetric matrices of dimension~$N \times N$. In both the
Grid and the Sum approach, the dimensionality of the matrix is
dependent on the box side length as it determines the size of the
discretized momentum space. The left panel of Fig.~\ref{fig:count}
shows the number of vectors
$\vec{q}\in\frac{2\pi}{L}\langle\zdrei\rangle$ with $|\vec{q}| <
\Lambda$, relevant for the Grid approach, as well as the number of
distinct absolute values of these vectors, relevant for the Sum
approach, for the two investigated cutoffs $\Lambda a=225$ and
$\Lambda a=314$. As $L$ increases, the vectors in the discrete space
$\frac{2\pi}{L}\langle\zdrei\rangle$ lie closer to each other,
eventually resembling the continuum. We therefore expect the total
number of points to be proportional to the volume of a sphere, scaling
as $(\Lambda L)^3$. This is indeed the case for $L > a/2$. For smaller
volumes, the behavior is slightly different because the discretized
space is still too ``coarse''. The number of distinct radii can also
be fitted to a power law, scaling like $(\Lambda L)^2$.

All results presented in this publication were obtained using a
sequential implementation. The runtimes range from seconds for the
smallest volumes to days for the larger volumes in the grid approach.
The length dependence of the runtime is depicted in the right panel of
Fig.~\ref{fig:count} for the two cutoffs and three different runs,
namely the Grid approach as well as the S-wave only and the full run
in the Sum approach. The runtime of the Strictly Perturbative approach
is almost identical to that of the S-wave run, while the runtime of
Partial Resummation approach is comparable to that of the Sum approach
including the G-wave. The scaling of the Sum approach is much better
than that of the Grid approach, as expected. In the latter, the
dimensionality of the matrix that has to be diagonalized in an
$O(N^3)$ operation scales like $(\Lambda L)^3$ compared to $(\Lambda
L)^2$ for the Sum approach. Therefore, we expect the runtime of the
Grid approach to scale with three additional powers of $\Lambda
L$. Indeed, fitting power laws to the runtime behavior yields a
scaling of $O\left((\Lambda L)^7\right)$ for the Sum approach and
$O\left((\Lambda L)^{10}\right)$ for the Grid approach. The inclusion
of the G-wave does not change the scaling of the code but merely
increases the runtime by about 10\%, independent of the cutoff. This
is a significant improvement in comparison to the Poisson approach
employed in previous publications, where the inclusion of the G-wave
increased the runtime by a factor of~4~\cite{Kreuzer:2009jp,phd}. By
using parallelized code, one should be able to significantly reduce
the runtime for larger volumes where the matrix dimensionality exceeds
about 2000. The volume range where this is the case depends on the
approach employed as well as the cutoff chosen.

\section{Summary and Outlook}
\label{summary}
We studied the volume dependence of three-boson states below the
three-body break-up threshold inside a cubic box with periodic
boundary conditions. A new framework has been developed that provides
access to the region above the boson-diboson break-up threshold and
allows for a significantly faster numerical implementation of the
higher partial waves. We derived an infinite set of coupled equations
for the partial waves of the boson-diboson amplitude. These equations
were solved for several cubic volumes with side lengths ranging from
$L=a$ to $L=a/4$. Proper renormalization of the results was explicitly
verified. We studied the effects from higher partial waves and found
them to be comparable to the variation due to the use of a finite
cutoff. We showed how these contributions can be treated
perturbatively, either by partial resummation or by the use of
perturbation theory for the eigenvalue equation. Both approaches yield
results that are in good agreement with the full calculation, with the
former performing slightly better and the latter offering a
significant reduction of runtime. This is an improvement compared to
our previous framework where higher partial waves play a more
pronounced role and their numerical treatment is more tedious.

The next step is to extend the framework to the three-nucleon system
and include higher orders of the EFT.  The latter is necessary in
order to perform precision extrapolations of finite volume
results. Work in this direction is in progress and will be the subject
of a future publication. With the new framework, it is also possible
to study states that extrapolate to a scattering state in the infinite
volume. This also allows one to examine, for example, three-body
effects in Lattice QCD simulations of nucleon-deuteron scattering. In
related work, it was shown that the compositeness of the deuteron
yields modifications of topological nature to L{\"u}scher's
formula~\cite{Bour:2011ef}.

We note that an extension of the L{\"u}scher formula relating the
infinite volume scattering phase shifts to the discrete energy levels
in a finite volume \cite{Luscher:1990ux} is implicitly contained in
our work. It provides the framework to determine the low-energy
constants of pionless EFT in the two- and three-body sector from
discrete energy levels in a cubic box. After this has been done, the
infinite-volume scattering observables can be calculated in pionless
EFT.

In summary, our results demonstrate that the finite volume corrections
for systems of three identical bosons are calculable and under
control. The role of higher partial waves was assessed and their
effect was shown to be of highly perturbative nature. With high
statistics Lattice QCD simulations of three-baryon systems within
reach~\cite{Beane:2009gs,Beane:2010em}, the calculation of the
structure and reactions of light nuclei appears feasible in the
intermediate future.

\begin{acknowledgments}
  We thank Hans-Werner Hammer for valuable discussions and for
  providing computational infrastructure.  A large part of the
  calculations has been performed on the CPU cluster at the HISKP,
  University of Bonn.  This work was supported in part by the National
  Science Foundation under CAREER award PHY-0645498, by the
  US-Department of Energy under contract DE-FG02-95ER-40907, and by
  University Facilitating Funds of the George Washington University.
\end{acknowledgments}

\appendix

\section{Derivation of Eq.~\eqref{eq:ipq}}
\label{app:ipq}
The partial wave components of the one-boson exchange part of the
interaction kernel~$\mathcal{Z}(E;p,q)$ are given by the quantity
$I_p^{(\ell)}(\vec{q})$ defined in Eq.~\eqref{eq:ipqdef},
\begin{equation*}
  I_p^{(\ell)}(\vec{q}) = \int \frac{\dd^2\hat{p}}{4\pi}\,
    K_{A_1 \ell}(\hat{p})\left[\sum_{C\in O}
      \left(\frac{1}{\alpha + \vec{p}\cdot C\vec{q}} + 
        \frac{1}{\alpha - \vec{p}\cdot C\vec{q}}\right)\right].
\end{equation*}
The integration can be carried out by expanding the kubic harmonic in
terms of spherical harmonics, transforming the integration variables
by a rotation~$R$ defined by $qR\hat{e}_z=C\vec{q}$ and using Wigner
D-matrices:
\begin{equation}
  \label{eq:ipq1}
  \begin{split}
    I_p^{(\ell)}(\vec{q}) &= \sum_{C\in O}\sum_m C_{\ell m}
      \int\frac{\dd^2\hat{p}}{4\pi} 
        \left(\frac{Y_{\ell m}(\hat{p})}{\alpha + \vec{p}\cdot C\vec{q}} + 
        \frac{Y_{\ell m}(\hat{p})}{\alpha - \vec{p}\cdot C\vec{q}}\right) \\
      &= \sum_{C\in O}\sum_m C_{\ell m}
      \int\frac{\dd^2\hat{p}}{4\pi} 
        \left(\frac{Y_{\ell m}(R\hat{p})}{\alpha + R\vec{p}\cdot C\vec{q}} + 
        \frac{Y_{\ell m}(R\hat{p})}{\alpha - R\vec{p}\cdot C\vec{q}}\right) \\
      &= \sum_{C\in O}\sum_m C_{\ell m} \sum_{\mp} D_{\mp m}^{(\ell)}(R)
      \int_{-1}^{+1}\int_0^{2\pi}\frac{\dd\cos\theta\,\dd\phi}{4\pi} 
        \left(\frac{Y_{\ell \mp}(\hat{p})}{\alpha + pq\cos\theta} + 
        \frac{Y_{\ell \mp}(\hat{p})}{\alpha - pq\cos\theta}\right)\\
      &= \sum_{C\in O}\sum_m C_{\ell m} \sum_{\mp} D_{\mp m}^{(\ell)}(R)
      N_\ell\delta_{\mp 0}
      \int_{-1}^{+1}\frac{\dd\cos\theta}{2} 
        \left(\frac{P_{\ell}(\cos\theta)}{\alpha + pq\cos\theta} + 
        \frac{P_{\ell}(\cos\theta)}{\alpha - pq\cos\theta}\right)\\
      &= \sum_{C\in O}\sum_m C_{\ell m} D_{0 m}^{(\ell)}(R)
      N_\ell\int_{-1}^{+1}{\dd\cos\theta}
      \frac{P_{\ell}(\cos\theta)}{\alpha - pq\cos\theta}\\
      &= \sum_{C\in O}\sum_m C_{\ell m} \frac{1}{N_\ell}
      Y_{\ell m}(C\hat{q})N_\ell\frac{2}{pq}
      Q_\ell\left(\frac{\alpha}{pq}\right)\\
      &= \frac{2}{pq}Q_\ell\left(\frac{\alpha}{pq}\right)
      \sum_{C\in O}K_{A_1 \ell}(C\hat{q})\\
      &= \frac{48}{pq}Q_\ell\left(\frac{\alpha}{pq}\right)K_{A_1 \ell}(\hat{q}).
  \end{split}
\end{equation}
Here, $N_\ell=\sqrt{\frac{2\ell+1}{4\pi}}$ is the normalization
constant of $Y_{\ell 0}$. The Wigner D-matrix is given by
$D_{0m}^{(\ell)}(\beta,\gamma,0) = \frac{1}{N_\ell}Y_{\ell
  m}(\gamma,\beta)$, where $\beta,\gamma$ are the Euler angles of the
rotation under consideration, which are by the definition of $R$ just
given by $C\hat{q}$. Further, we use a well-known integral
representation of the Legendre functions of the second
kind~$Q_\ell$~\cite{Abramowitz} and, in the final step, the invariance
of the $K_{A_1 \ell}$ under cubic rotations.


\end{document}